\documentclass{article}
\usepackage{spconf,amsmath,graphicx, multirow, dblfloatfix}
\usepackage{adjustbox,booktabs}
\usepackage[hyphens]{url}
\usepackage{soul,color}
\usepackage{amssymb}
\usepackage{standalone}

\usepackage{flushend}

\title{Convolutional Recurrent Neural Networks \\for Bird Audio Detection}
\name{Emre~\c{C}ak{\i}r, Sharath Adavanne, Giambattista Parascandolo, Konstantinos Drossos, Tuomas Virtanen\thanks{The research leading to these results has received funding from the European Research Council under the European Union’s H2020 Framework Programme through ERC Grant Agreement 637422 EVERYSOUND. The authors wish to acknowledge CSC-IT Center for Science, Finland, for computational resources.}}
\address{Department of Signal Processing, Tampere University of Technology}

\begin{document}
%
\maketitle
\begin{abstract}
Bird sounds possess distinctive spectral structure which may exhibit small shifts in spectrum depending on the bird species and environmental conditions. In this paper, we propose using convolutional recurrent neural networks on the task of automated bird audio detection in real-life environments. In the proposed method, convolutional layers extract high dimensional, local frequency shift invariant features, while recurrent layers capture longer term dependencies between the features extracted from short time frames. This method achieves 88.5\% Area Under ROC Curve (AUC) score on the unseen evaluation data and obtains the second place in the Bird Audio Detection challenge.
\end{abstract}
\section{Introduction} 
\label{sec:intro}
Bird audio detection (BAD) is defined as identifying the presence of bird sounds in a given audio recording. In many conventional, remote wildlife-monitoring projects, the monitoring/detection process is not fully automated and requires heavy manual labor to label the obtained data (e.g. by employing video or audio)~\cite{buxton2012measuring, marques2013estimating}. In certain cases such as dense forests and low illumination, automated detection of birds in wildlife can be more effective through their sounds compared to visual cues. Besides, acoustic monitoring devices can be easily deployed to cover wide ranges of land. This indicates the need for automated BAD systems in various aspects of biological monitoring. For instance, it can be applied in the automatic monitoring of biodiversity, migration patterns, and bird population densities~\cite{marques2013estimating, borker2014vocal}. BAD systems can be augmented with another classifier to determine the species of the detected birds~\cite{graciarena2011bird}. Using an automated BAD system as preprocessing/filtering step to determine the bird species would be beneficial especially for remote acoustic monitoring projects, where large amount of audio data is employed.


In this regard, the Bird Audio Detection challenge~\cite{stowell2016bird} is organized with an objective to stimulate the research on BAD systems which can work on real life bioacoustics monitoring projects. The challenge provides three bird audio datasets recorded in different acoustic environments. Two of the datasets are provided with bird call annotations to be used as development data. The final dataset consists of recordings from a different physical environment and it is employed as the evaluation data. An extensive review on the recent work on BAD can also be found in~\cite{stowell2016bird}.

Bird sounds can be broadly categorized as vocal and non-vocal sounds (such as bill clattering, and drumming of woodpeckers)~\cite{howell1995guide}. Since non-vocal bird sounds are harder to be associated with birds without any visual cues, the research on BAD has been mostly focused on vocal sounds, as in this work. Vocal sounds can be further categorized as bird calls and bird songs. Bird calls are often short and serve a particular function such as alarming or keeping the flock in contact. Bird songs are typically longer and more complex than bird calls, and they often possess temporal structure which are melodious to human ears~\cite{ehrlich}. Mating calls can be given as example to bird songs. Vocal bird sounds include distinctive spectral content often including harmonics. Alarm calls tend to be high-pitched with rapid modulations (to get maximum attention), whereas lower frequency calls are common in densely vegetated areas to avoid signal degradation due to reverberation~\cite{derryberry2009ecology}. Furthermore, depending on the environmental conditions (e.g. ambient noise level, vegetation density) and the bird species, bird sounds may exhibit certain local frequency shift variations~\cite{derryberry2009ecology}. Therefore, a BAD system should be able to capture melodic cues in time domain, and also should be robust to local frequency shifts. 

\begin{figure}
  \centering
  \includegraphics[width=0.95\columnwidth]{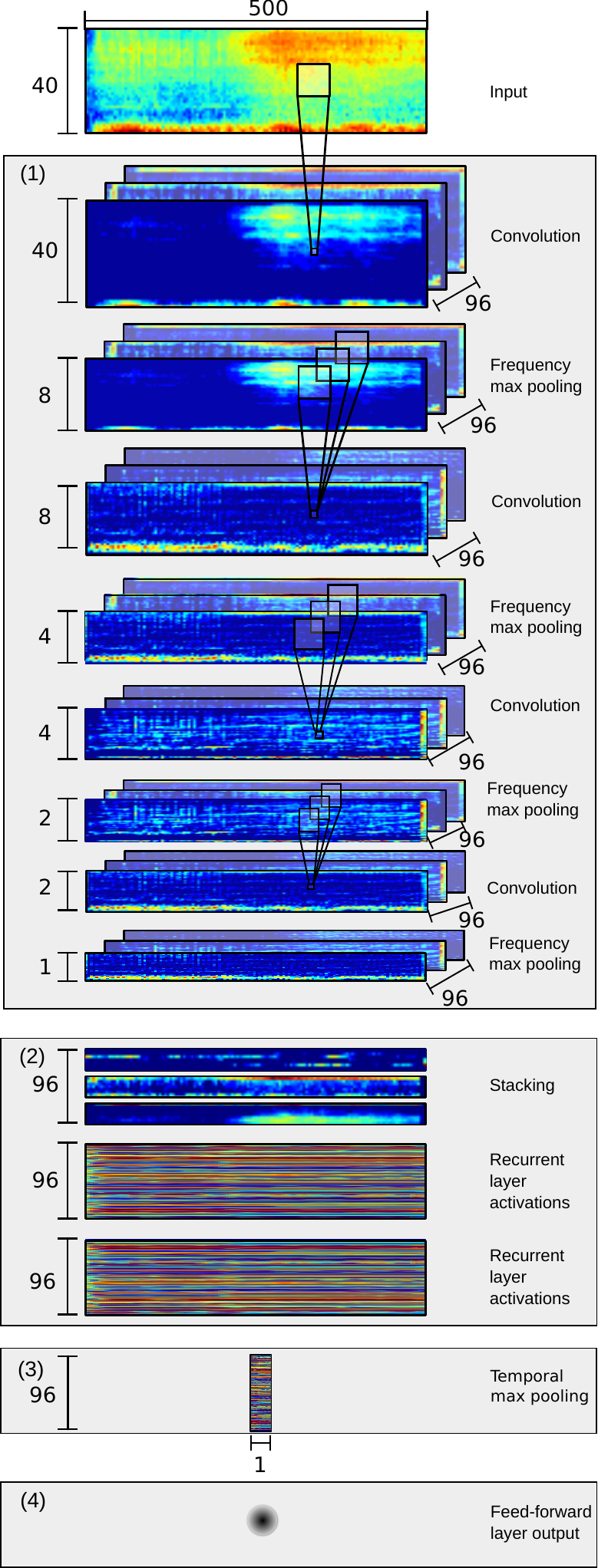}
  \caption{Illustration of the CRNN architecture proposed for bird audio detection.}
  \label{fig:crnn}
\end{figure}

Convolutional neural networks (CNN) are able to extract higher level features that are invariant to local spectral and temporal shifts. Recurrent neural networks (RNNs) are powerful in learning the longer term temporal context in the audio signals. In this work, we combine these two approaches in a convolutional recurrent neural network (CRNN) and apply it over spectral acoustic features for the BAD challenge. This method consists of slight modification (temporal max-pooling to obtain file-level estimation instead of frame-level estimation) and hyperparameter fine-tuning for the challenge over the CRNN proposed in~\cite{emre_TASLP2016}, where it has provided state-of-the-art results on various polyphonic sound event detection and audio tagging tasks. Similar approaches combining CNNs and RNNs have been presented recently in ASR~\cite{sainath2015convolutional} and music classification~\cite{choi2016convolutional}.

The rest of the paper is organized as follows. The employed acoustic features and the proposed CRNN for the BAD are presented in Section~\ref{sec:method}. Dataset settings, metrics, and method configuration are reported in Section~\ref{sec:eval}. In Section~\ref{sec:results} are the results and their discussion, followed by the conclusions in Section~\ref{sec:concl}. 

\section{Method}\label{sec:method}
The proposed method consists of two stages. In the first stage, spectro-temporal features (spectrogram) are extracted from the raw audio recordings to be used as the sound representation. In the second stage, a CRNN is used to map the acoustic features to a binary estimate of bird song presence. CRNN parameters are obtained by supervised learning using material that consists of acoustic features extracted from a training database and the annotations of bird song activity.


\subsection{Features} 
\label{sec:feat}
The utilized spectro-temporal features are log mel-band energies, extracted from short frames. These features has been shown to perform well in various audio tagging and sound event detection tasks~\cite{dcase2016task3web, emre2015, emre_TASLP2016}. First, we obtained the magnitude spectrum of the audio signals by using short-time Fourier transform (STFT) over 40 ms audio frames of 50\% overlap, windowed with Hamming window. Duration of each audio file in the challenge dataset is 10 seconds, resulting to 500 frames for each file. Then, 40 log mel-band energy features were extracted from the magnitude spectrum. Librosa library~\cite{mcfee2015librosa} was used in the feature extraction process.

Keeping in mind that bird sounds are often contained in a relatively small portion of the frequency range (mostly around 2-8 kHz), extracting features from that range seems like a good approach. However, experiments with features from the whole frequency range (from 0 Hz to Nyquist frequency) provided better results, and were therefore utilized in the proposed method.

\subsection{Convolutional recurrent neural networks}
\label{sec:crnn}
The CRNN proposed in this work, depicted in Figure~\ref{fig:crnn}, consists of four parts: 

\begin{enumerate}
	\item{convolutional layers with rectified linear unit (ReLU) activations and non-overlapping pooling over frequency axis}
    \item{gated recurrent unit (GRU)~\cite{cho2014properties} layers}
    \item{a temporal max-pooling layer, and}
    \item{a single feedforward layer with a single unit and sigmoid activation, as the classification layer.}
\end{enumerate}

A time-frequency representation of the data is fed to the convolutional layers and the activations from the filters of the last convolutional layer are stacked over the frequency axis and fed to the first GRU layer. The extracted representations over each time frame (from the last GRU layer) are used as input to the temporal max-pooling layer. Output of the max-pooling layer is employed as input to the classification layer. Output of the classification layer is treated as the bird audio probability for the audio file. The aim of the network learning is to get the estimated bird audio probabilities as close as to their binary target outputs, where target output is 1 if any bird sound is present in a given recording, and 0 vice versa.

The network is trained with back-propagation through time using Adam optimizer~\cite{adamKeras} and binary cross-entropy as the loss function. In order to reduce overfitting of the model, early stopping was used to stop training if the validation data AUC score did not improve for 50 epochs. For regularization, batch normalization~\cite{batchNorm} was employed in convolutional layers and dropout~\cite{Dropout} with rate 0.25 was employed in convolutional and recurrent layers. Keras deep learning library~\cite{chollet2015keras} has been used to implement the network.

The proposed method differs from our other submission~\cite{adavanne2017bad} for the challenge (which came in fifth place) in the following ways: we use a single set of acoustic features, smaller max pool size in frequency domain and no max pooling in time domain, no maxout for the output layer, and the whole method consists of a single branch with unidirectional GRU. In addition, considering the auxiliary data augmentation and domain adaptation techniques applied in~\cite{adavanne2017bad}, the proposed method is less complex and still performs better in the given BAD challenge.

\section{Evaluation} 
\label{sec:eval}
\subsection{Datasets}
\label{ssec:data}
The Bird Audio Detection challenge~\cite{stowell2016bird} consists of a development and an evaluation set. The development set consists of \textit{freefield1010} (field recordings gathered by the \footnote{http://freesound.org/}{FreeSound} project) and \textit{warblr} (crowd-sourced recordings collected through smartphone app) datasets, and the evaluation set consists of \textit{chernobyl} (collected by unattended recorders in Chernobyl exclusion zone) dataset. Recordings in all the datasets are around 10 seconds long, single channel, and sampled at 44.1 kHz. The annotations for the recordings are binary - bird calls present or absent. The total duration of the available recordings is approximately 68 hours, which makes the dataset a valuable source for detection methods that require large amount of material. The statistics of the datasets are presented in Table~\ref{Table:1}. 


From the development set, we create five different splits with 60\% training, 20\% validation, and 20\% testing set distribution. Each split has an equal distribution of birds call present and absent, i.e. 60\% of all the development data with present bird call annotation is included in training data, and the same is valid for absent bird call annotations. Different splits are obtained by randomly shuffling the recordings list and re-partitioning the data in given proportions. All development set results are the average performance over the splits. For the challenge submission, the CRNN is trained on single split of 80\% training and 20\% validation done on development set, with equal distribution of classes. 

\begin{table}
\centering
\caption{Bird audio detection challenge \cite{stowell2016bird} dataset statistics}
\label{Table:1}
\begin{tabular}{l|lll}
\multirow{2}{*}{Dataset} & \multicolumn{3}{c}{Bird call} \\\cline{2-4}
                         & Present        & Absent       & Total\\\hline
freefield1010            & 5755           & 1935         & 7690\\
warblr                   & 1955           & 6045         & 8000\\
chernobyl				 & ?			  & ?			 & 8620\\\hline
Total                    & 7710 + ?           & 7980 + ?         & 24310
\end{tabular}
\end{table}

\vspace{-4mm}
\subsection{Evaluation Metric and Configuration}
\label{ssec:configuration}

The BAD system output is evaluated from the receiver operating characteristic (ROC) using the AUC measurement. 

In order to obtain the optimal hyperparameters for the given task, we run a hyperparameter grid search over the validation set. The grid search covers each of the combinations of the following hyperparameter values: the number of CNN feature maps/RNN hidden units (the same amount for both) \{96, 256\}; the number of recurrent layers \{1, 2, 3\}; and the number of convolutional layers \{1, 2, 3 ,4\} with the following frequency max pooling arrangements after each convolutional layer \{(4), (2, 2), (4, 2), (8, 5), (2, 2, 2), (5, 4, 2), (2, 2, 2, 1), (5, 2, 2, 2)\}. Here, the numbers denote the number of frequency bands at each max pooling step; e.g., the configuration (5, 4, 2) pools the original 40 bands to one band in three stages: 40 bands $\rightarrow$ 8 bands $\rightarrow$ 2 bands $\rightarrow$ 1 band. The final network configuration is given in Table~\ref{tab:params}.

\begin{figure*}[!ht]
\centering
\includegraphics[width=.63\textwidth]{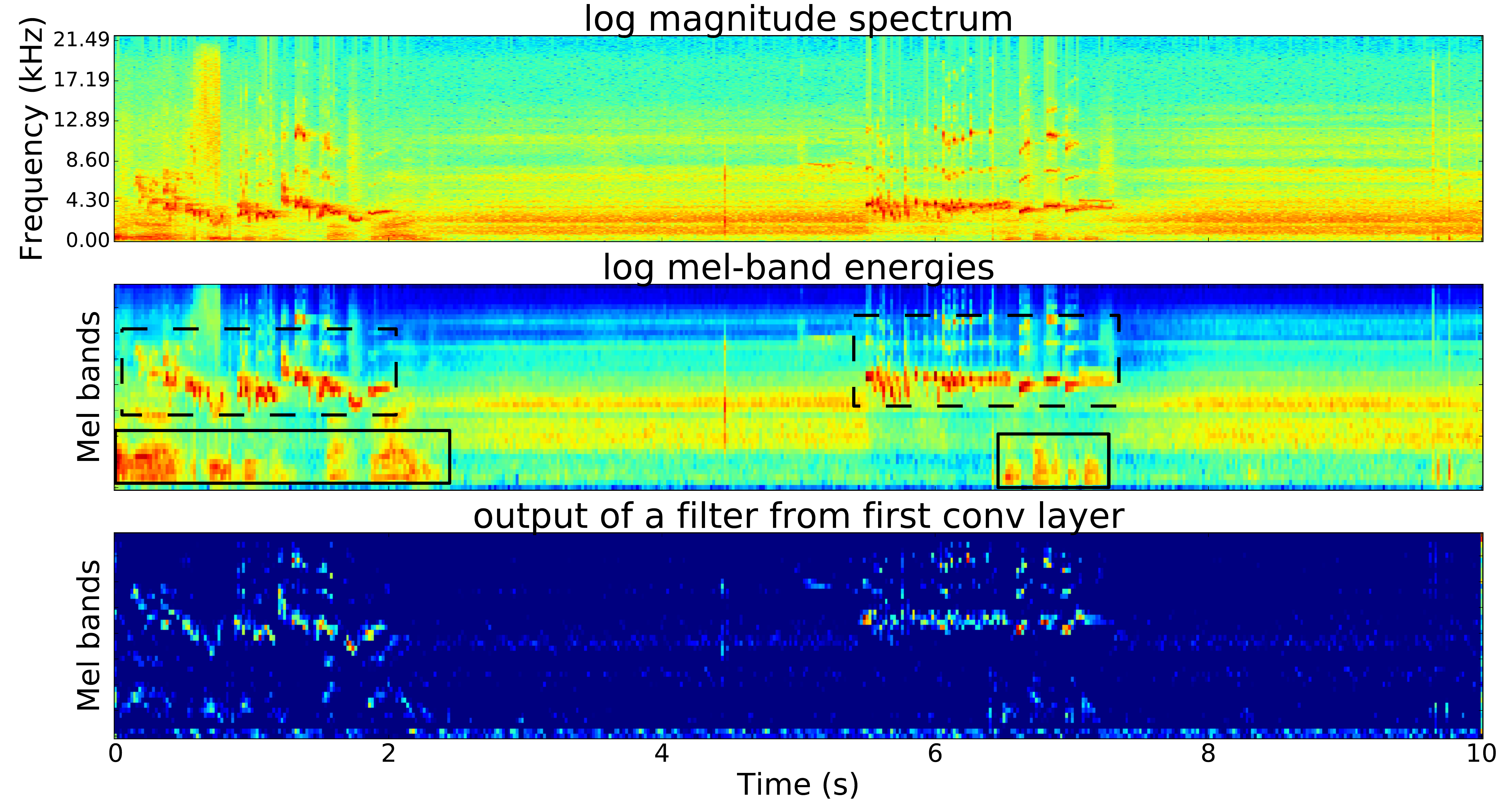}
\caption{Log magnitude spectrum (top), log mel-band energies (middle) and a single filter output from first convolutional layer (bottom) for \textit{000a3cad-ef99-4e5e-9845.wav}. Dashed boxes mark the components due to bird sounds, and solid boxes mark the components due to two people speaking.}
\label{fig:result_ex}
\end{figure*}


\begin{table}[t]
\centering
\caption{Final hyperparameters used for the evaluation based on the validation results from the hyperparameter grid search.}
\label{tab:params}
\begin{tabular}{ @{} l| *{1}{c c c|}}
\hline
& Hyperparameters \\
\hline
\# convolutional layers					&4	\\
Filter shape					&5-by-5\\
pool size 						&(5,2,2,2) \\
\# recurrent layers					&2	\\
\# feature maps/hidden units 	&96\\
\# Parameters					&806K\\
\end{tabular}
\end{table}

\subsection{Baseline}
In this work, we trained a CNN to be used as a baseline and also to understand the benefit of using recurrent layers after the convolutional layers. Based on the information given after the challenge, most of the submissions also use CNN as their classifier, and therefore it can be deemed as an appropriate baseline for the proposed method. The optimal parameters for CNN is found with a similar grid search as explained in Section~\ref{ssec:configuration}, the only difference is that we replace the recurrent layers with feedforward layers. Each feedforward layer had shared weights between timesteps.

For comparison, we also provide the scores from the top three submissions for the challenge. Both methods use CNN as classifier (therefore labeled as \textit{CNN2}~\cite{grill} and \textit{CNN3}~\cite{topel}), they use mel spectrogram as input features, and they apply frequency and time shift as data augmentation techniques. Both methods apply pseudo-labeling (i.e. including the very confident detections from the test set into training set) and they further apply model ensembling over the networks.

\section{Results and discussion}
\label{sec:results}
AUC scores for the baseline CNN and the proposed CRNN methods on development and evaluation sets are presented in Table~\ref{Table:2}. AUC for development set is obtained from the mean test AUC of the five splits. Although the performance difference between CNN and CRNN is minimal for the development data, CRNN performs significantly better for the evaluation data. Considering that the evaluation data includes recordings from different environmental and recording conditions than the development data, one can say that CRNN does a better job of generalizing over bird sounds in different conditions. For both methods, the validation data AUC score reaches to about 92\% in the very first epoch and reaches its peak in about 20 epochs. To compare with the other top submissions, CNN2 reaches 88.7\% AUC and CNN3 obtains 88.2\% on the evaluation data.

In order to provide some insight on the features and network outputs, one of the recordings from the evaluation set (namely \textit{000a3cad-ef99-4e5e-9845.wav}) has been specifically investigated. The top panel represents the magnitude spectrum (in log scale) for the recording, the middle panel shows the normalized log mel band energies which are used as input for the network, and the bottom panel represents the output from one of the filters in the first convolutional layer before max-pooling. When we compare the top two panels, we notice that with log mel band energies, the frequency components due to speech and bird sounds become very distinguishable. In addition, by looking at the filter outputs in the bottom panel, one can say that this filter has learned to react to the bird sound components and mostly ignore the rest for the given audio recording. The trained CRNN outputs a probability of 94.7\% for a bird sound in this recording.   

\begin{table}
\centering
\caption{AUC scores on development and evaluation sets}
\label{Table:2}
\begin{tabular}{l|cc}
 
Dataset &  			\multicolumn{2}{c}{Method}  \\\cline{2-3}
					& CNN	& CRNN \\\hline
Development     	& 95.3		& \textbf{95.7} \\
Evaluation			& 85.5		& \textbf{88.5}		
\end{tabular}%
\end{table}

Since the amount of available material is quite large (about 68 hours), we did not further experiment on various data augmentation techniques. 
For the challenge submission, we experimented with a model ensemble method: 11 networks with the same architecture and different initial random weights (obtained by sampling from different random seeds) were trained and the estimated probabilities from each network were averaged to obtain the ensemble output. Although this method improved the prior AUC results (calculated from a small portion of the evaluation data) from 88.3 to 89.4, it performed worse in the final results (88.2 vs. 88.5). The authors do not have a clear reasoning for this contradiction. 


\section{Conclusion} 
\label{sec:concl}

In this work, we propose using convolutional recurrent neural networks for bird audio detection as a part of a research challenge. The proposed method shows robustness for the local frequency shifts and is able to utilize longer term temporal information. Both of these features are essential for a generalized, context independent BAD system. The method achieves 88.5\% AUC score and obtains the second place in Bird Audio Detection challenge.

\bibliographystyle{IEEEtran}
\bibliography{refs}

\end{document}